\documentclass[english,aps,pra,superscriptaddress,reprint,twocolumns]{revtex4-1}
\usepackage[T1]{fontenc}
\usepackage{amsmath,amssymb,graphicx,mathrsfs,xcolor,babel,textcomp,esint}
\usepackage[unicode=true,pdfusetitle,bookmarks=true,bookmarksnumbered=false,bookmarksopen=false,breaklinks=true,pdfborder={0 0 0},backref=false,colorlinks=true]{hyperref}

\begin{document}

\title{Terahertz Emisssion from Quantum Interference of Electron Trajectories}

\author{Yizhu Zhang}
\affiliation{Center for Terahertz Waves and College of Precision Instrument and Optoelectronics Engineering, Key Laboratory of Opto-electronics Information and Technical Science, Ministry of Education, Tianjin University, China}
\affiliation{Shanghai Advanced Research Institute, Chinese Academy of Sciences, Shanghai 201210, China}
\author{Kaixuan Zhang}
\affiliation{Shanghai Advanced Research Institute, Chinese Academy of Sciences, Shanghai 201210, China}
\affiliation{University of Chinese Academy of Sciences, Beijing 100049, China}
\author{Tian-Min Yan}
\email{yantm@sari.ac.cn}
\affiliation{Shanghai Advanced Research Institute, Chinese Academy of Sciences, Shanghai 201210, China}
\author{Y. H. Jiang}
\email{jiangyh@sari.ac.cn}
\affiliation{Shanghai Advanced Research Institute, Chinese Academy of Sciences, Shanghai 201210, China}
\affiliation{University of Chinese Academy of Sciences, Beijing 100049, China}
\affiliation{ShanghaiTech University, Shanghai 201210, China}

\begin{abstract}
The semiclassical electron trajectory, the so-called quantum orbits, is employed to explain the terahertz wave generation (TWG) in dual-color strong field, and the feasibility of the theory is validated by the measurement. We find that TWG stems from quantum path interference of partial electron wavepacket released at the neighbouring cycles of the dual-color electric field, manifesting the temporal Young's double-slit interference of single electron. The trajectories released from neighbouring cycles, creating TWG, also account for intercycle interference fringes in the photoelectron momentum distribution, whereas no signature of TWG is found when taking trajectories only from a single cycle.

\end{abstract}
\maketitle

The semiclassical theory is a general and widespread method in the strong-field physics, in which the electron tunnelling from the barrier, formed by both the Coulombic and strong laser field, is treated quantum mechanically, and its subsequent motion is envisioned as the Newtonian particle trajectory, primarily consisting in the oscillation of free electron in the laser field. This approach provides an intuitive and semiclassical picture of the electron dynamics, which is hardly accessible with fully quantum-mechanical calculation, simultaneously including effects of quantum tunneling, propagation and interference of electron wave packet. The greatest success of the semiclassical theory is to explain the characteristics of above-threshold ionization (ATI) \cite{Corkum1989} and high-harmonic generation (HHG) \cite{Corkum1994,Lewenstein1994}. In semiclassical theory, the quantum interference of electron trajectories released from the neighbouring cycles results in the electron energy with discrete ATI peaks spaced by the laser photon energy. Similarly, the electron trajectories from neighbouring peaks recollide into parent ions to emit photon wavepackets, which interfere with each other, creating odd-order HHGs. Derived from the semiclassical treatment, the quantum trajectory method (QTM) has further elucidated the subtle features in photoelectron momentum distributions (PEMDs), e.g. low-energy feature \cite{Yan2010} and holographic structure \cite{Huismans2011}. The electron correlation in the double ionization has been successfully illuminated with the QTM in a straightforward and intuitive way \cite{Haan2006,Ye2008}, and recently, the QTM has been extended to understand HHG in solids \cite{Li2019}.

The terahertz (THz) wave generation (TWG) using dual-color ($\omega-2\omega$) field in the gas-phase medium, since been firstly observed \cite{Cook:00}, is the most widely used method for table-top generation of the ultra-broadband strong THz pulse, and the mechanism has been subjected to the continuous interrogation theoretically and experimentally \cite{Kim2007,Andreeva2016,Zhang2017}. The TWG in the dual-color field was initially explained by the four-wave mixing model \cite{Cook:00}, as in crystal nonlinear optics and nonlinear spectroscopy. Later, the photocurrent (PC) model was developed to attribute TWG to the time-variant plasma current \cite{Kim2007}. Simultaneously, much effort has been made to explain the TWG within the strong-field theoretical framework, including the strong-field approximation (SFA) model \cite{Zhou2009} and numerical solving time-dependent Schr\"{o}dinger equation \cite{Karpowicz2009}. Recently, our experimental and theoretical works indicated that the TWG originates from the continuum electron under single active electron approximation, and established the link between quantum mechanical SFA and classical PC models \cite{Zhang2020a,Zhang2020b}. However, which kind of electron behaviors in convoluted tunneling dynamics produces the TWG still remains indistinct. 

\begin{figure}[h]
  \includegraphics[width=\linewidth]{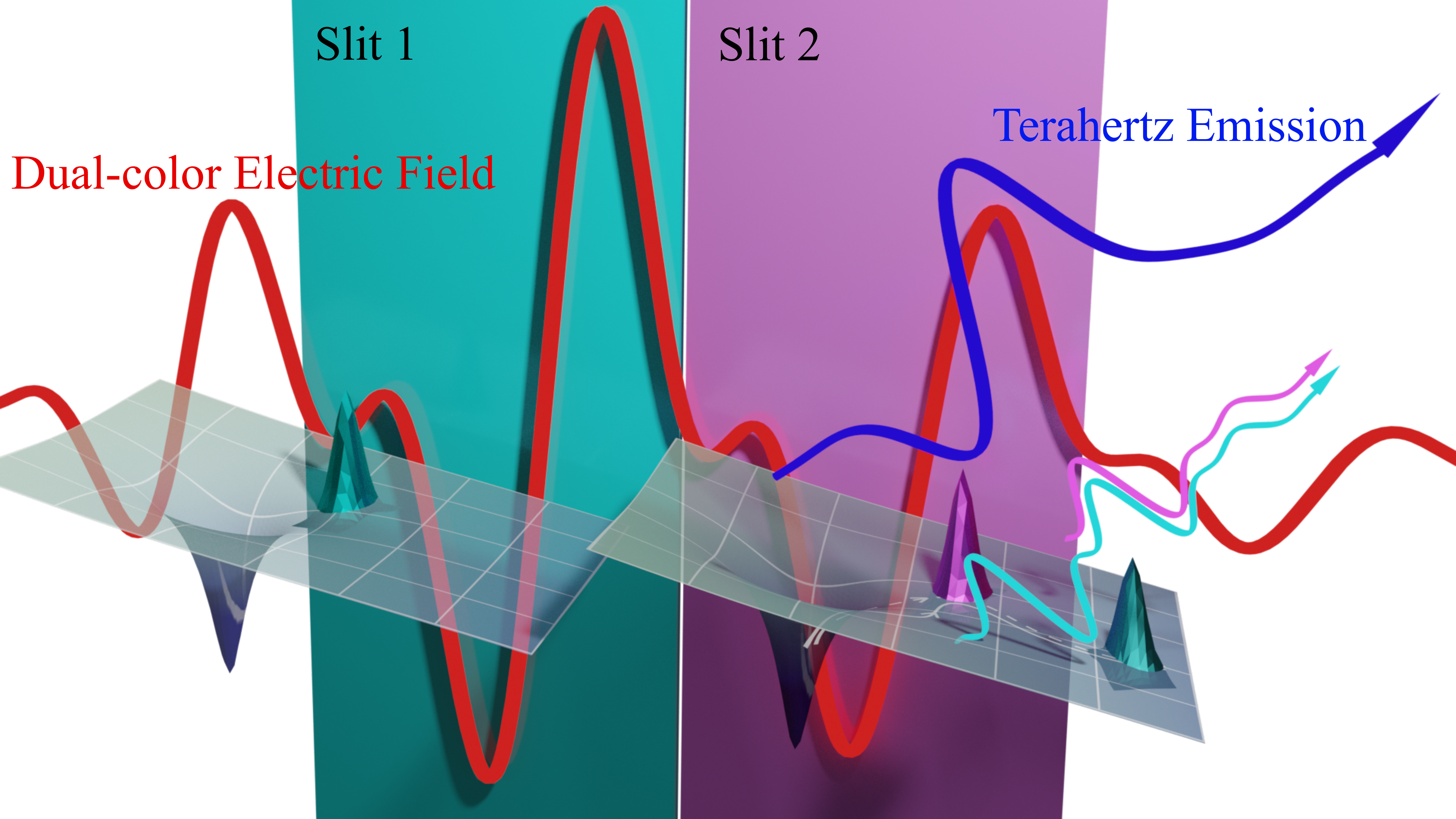}
  \caption{Illustrative schematics of the quantum trajectory explanation for terahertz emission. The dual-color electric field (red curve) excites the atoms and terahertz radiation (blue curve) emits. The electron wavepacket (cyan wavepacket) is partially released from the distorted atomic potential within the temporal window ``slit 1'' (cyan region), followed by the subsequent release of electron wavepacket (magenta wavepacket) within the temporal ``slit 2'' (magenta region). Two separately ionized continuum electron wavepackets are spreading and propagating in the dual-color laser, that can be described as an ensemble of quantum trajectories. Each quantum trajectory produces secondary radiation independently, and the emitted photon wavepackets are shown as the cyan and magenta curves. The photon wavepackets appearing within different ionization slits temporally interfere, leading to the constructive interference in terahertz frequency domain.}
\end{figure}

In this letter, the QTM is employed to investigate the TWG in dual-color laser field, and the feasibility is experimentally presented in parallel with the theory. Here, we provide a new viewpoint for the origin of the TWG from the quantum trajectory perspective (Fig. 1): The THz emission comes from the interference between the continuum electron wavepacket of the single electron released at the neighbouring cycles of the dual-color field. The TWG phenomenon, interpreted in terms of interfering quantum trajectories appearing at different temporal ``slits'', provides a textbook example for the quantum Young's ``double-slit'' experiment in time domain. The link between the TWGs and PEMDs indicates that the intra-cycle interference of electron trajectories creates both the TWG and ``ring'' pattern fringes in PEMDs; the inter-cycle interference of trajectories, though creating the ``wing'' fringes in PEMDs, yet cannot produce the terahertz pulse. The TWG shares the similar physical picture with the discrete peaks in ATI and HHG resulting from single-electron interference. Present studies extend the application landscape of the QTM and unify the TWG with other strong-field phenomena from the semiclassical theory.

The QTM is derived from the SFA model. In its simplest form, only the electron from direct ionization is taken into account, and the recollision of electron back to the parent ion is not considered. The bound electron is released at the ionization instant $t_{\mathrm{ion}}$ and subsequently driven only by the strong laser field without interaction with the parent ion anymore. The final state of the electron wavefunction with asymptotic momentum $\boldsymbol{p}$ gives the PEMDs with the SFA probability amplitude $M_{\boldsymbol{p}}^{(\mathrm{SFA})}=-i\int_{0}^{\infty}\langle\Psi_{\boldsymbol{p}}^{(\mathrm{GV})}(t_{\mathrm{ion}})\mid zE(t_{\mathrm{ion}})\mid\Psi_{0}(t_{\mathrm{ion}})\rangle dt_{\mathrm{ion}}.$ $\mid\Psi_{0}(t_{\mathrm{ion}})\rangle$ is the electron wavefunction at $t_{\mathrm{ion}}$, and $E(t)$ is the strong laser field. The continuum electron wavefunction in $\boldsymbol{p}$ space, drifted by the vector potential $\boldsymbol{A}(t)=-\int_{-\infty}^{t}\boldsymbol{E}(t')dt'$, can be approximated with the Gordon-Volkov state $\mid\Psi_{\boldsymbol{p}}^{(\mathrm{GV})}(t_{\mathrm{ion}})\rangle=e^{-iS_{\boldsymbol{p}}(t_{\mathrm{ion}})}\mid\boldsymbol{p}+\boldsymbol{A}(t_{\mathrm{ion}})\rangle$, where $S_{\boldsymbol{p}}(t_{\mathrm{ion}})=\int_{t_{\mathrm{ion}}}^{\infty}\frac{1}{2}[\boldsymbol{p}+\boldsymbol{A}(t')]^{2}dt'$. 

The QTM is introduced using the saddle-point approximation to numerically solve the time integral of $M_{\boldsymbol{p}}^{(\mathrm{SFA})}$. The $M_{\boldsymbol{p}}^{(\mathrm{SFA})}$ can be approximated by a sum over the stationary contributions, leading to $\overset{\backsim}{M}_{\boldsymbol{p}}^{(\mathrm{SFA})}\simeq\underset{\alpha}{\sum}C_{\boldsymbol{p}}^{(\alpha)}e^{-iS_{\boldsymbol{p}I_{p}}^{(\alpha)}}$, where $S_{\boldsymbol{p}I_{p}}^{(\alpha)}=S_{\boldsymbol{p}I_{p}}(t_{\boldsymbol{p}}^{(\alpha)})$, $S_{\boldsymbol{p}I_{p}}(t_{\boldsymbol{p}}^{(\alpha)})=\int_{t_{\boldsymbol{p}}^{(\alpha)}}^{\infty}\left\{ \frac{1}{2}\left[\boldsymbol{p}+\boldsymbol{A}(t')\right]^{2}+I_{p}\right\} dt'$, and $C_{\boldsymbol{p}}^{(\alpha)}$ is a prefactor. $I_{\boldsymbol{p}}$ is the ionization potential of the hydrogen atom. The $\alpha$th saddle point $t_{\boldsymbol{p}}^{(\alpha)}$ fulfills the stationary action equation $\frac{\partial S_{\boldsymbol{p}I_{p}}}{\partial t}|_{t_{\boldsymbol{p}}^{(\alpha)}}=0$, and the solution $t_{\boldsymbol{p}}^{(\alpha)}$ is complex. The saddle-point equations define a number of quantum orbits that are analogous to classical orbits, but propagating in the complex plane. The $S_{\boldsymbol{p}I_{p}}$ is the action integral, which is the accumulation of the instantaneous kinetic energy of the electron along the trajectory. The PEMDs, shown in Fig. 2(b), are obtained with the coherent sum of $S_{\boldsymbol{p}I_{p}}^{(\alpha)}$ of quantum orbits.  

The motion of the electronic wave packet is described as series of quantum trajectories propagating in the complex-time plane. The propagation paths, defined by the saddle point $t_{\boldsymbol{p}}$, can be separated into the propagation trajectory along the imaginary axis from $t_{\boldsymbol{p}}$ down to the real axis $t_{0}=\mathrm{Re}t_{\boldsymbol{p}}$ and then along the real axis from $t_{0}$ to the observation time $t$. The trajectory $\boldsymbol{r}(t)$ can be solved according to the initial condition \cite{Yan2013}. The real-space trajectory $\boldsymbol{r}_{t>t_{0}}(t)$, propagating along the real-time axis $t>t_{0}$, is analogous to the classical Newton's motion of the free electron in the real-position space after the ionization instant, that reads $\boldsymbol{r}_{t>t_{0}}(t)=\int_{t_{0}}^{t}\left[\boldsymbol{A}(t')+\boldsymbol{p}\right]dt'+\boldsymbol{r}(t_{0})$, where the $\boldsymbol{r}(t_{0})$ is the tunnelling exit in the real-position space. 

The secondary radiation comes from the acceleration of the laser-assisted oscillation of the free electron, described as the classical Maxwell's electromagnetic theory. The secondary radiation is temporally overlapping of photon wavepackets induced by the quantum trajectories
\begin{equation}
\boldsymbol{E}_{\text{\ensuremath{\mathrm{Har}}}}(t)\sim\sum_{\alpha}w^{(\alpha)}\frac{\partial^{2}\boldsymbol{r}_{t>t_{0}}^{(\alpha)}(t)}{\partial t^{2}}\label{eq:Harmonic radiation}
\end{equation}
where $w^{(\alpha)}=|e^{-iS_{\boldsymbol{p}I_{p}}^{(\alpha)}}|^{2}$ is the weight factor, which physically describes the ionization possibility of the \textbf{$\alpha$}th electron trajectory. The THz radiation $\boldsymbol{E}_{\mathrm{THz}}$ 
is acquired by filtering out the corresponding Fourier harmonic components of $\boldsymbol{E}_{\mathrm{Har}}(t)$. The QTM for the TWG is a semiclassical method: The tunnelling ionization is described quantum mechanically; the weight of electron trajectories represents the ionization possibilities. The ensemble of classcial electron trajectories corresponds to the spreading and propagating electron wave packet. The electromagnetic radiation is anticipated by the classical Maxwell's theory with recourse to the Newton's trajectories of the free electron. 

\begin{figure}[h]
  \includegraphics[width=\linewidth]{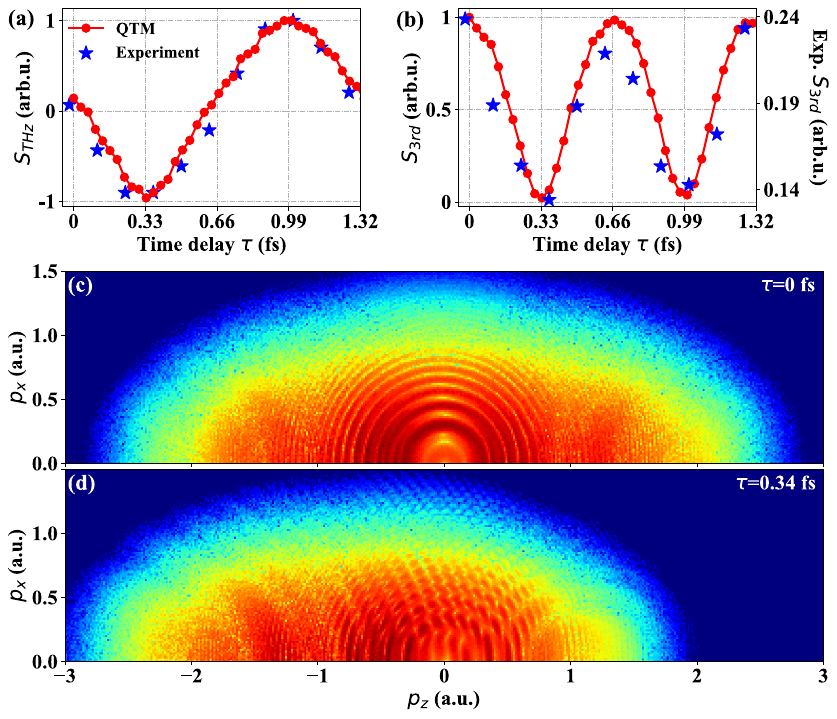}
  \caption{The $\tau$-dependent terahertz generation with quantum trajectory method. (a1) and (a2) The $S_{\mathrm{THz}}(\tau)$ and $S_{3\omega}(\tau)$ of the QTM (red dot) and experiment (blue star). (b1) and (b2) The logarithmically scaled $w(\boldsymbol{p},\tau=0 \textrm{ fs})$ and $w(\boldsymbol{p},\tau=0.34 \textrm{ fs})$ in the $p_{z}-p_{x}$ plane, predicted with the QTM. The polarization of the $\omega$ and $2\omega$ electric fields is aligned along the $p_{z}$ axis. In the figure, all the values are normalized to the maximum.}
\end{figure}

The QTM is experimentally validated with the \textit{sin}-type yield as a function of the  $\omega-2\omega$ phase delay $\tau$, which is the most robust characteristics of TWG. The THz peak-peak amplitudes $S_{\mathrm{THz}}(\tau)$, both the QTM (red dots) and measurement (blue stars), are shown in Fig. 2(a1) for comparison. The $S_{\mathrm{THz}}(\tau)$ is calculated using Eq. \ref{eq:Harmonic radiation}, for details described in \textit{Supplemental Material S.1}. The $S_{3\omega}(\tau)$, shown in Fig. 2(a2), are jointly measured with  $S_{\mathrm{THz}}(\tau)$ for experimentally \textit{in situ} calibration of the phase delay $\tau$ \cite{Zhang2020a}. The experimental detail is referred in \textit{Supplemental Material S.2}. The PEMDs $w(\boldsymbol{p})$ at $\tau=0 \textrm{ fs}$ and $\tau=0.34 \textrm{ fs}$, the minimum and maximum in $|S_{\mathrm{THz}}(\tau)|$  respectively, are shown in Fig. 2(b), which are calculated with the formula of $\overset{\backsim}{M}_{\boldsymbol{p}}^{(\mathrm{SFA})}$.  The sufficient sampling $N$ in the $\boldsymbol{p}$ space is necessary to visualize the subtle structures in the $w(\boldsymbol{p})$. However, only few quantum trajectories are necessary to reproduce the TWG results, and the harmonic spectra rapidly converge as the $N$ increasing. The QTM predicted $S_{\mathrm{THz}}(\tau)$ is consistent with commonly used theoretical models including the continuum-continuum transition in SFA and PC model \cite{Zhang2020b}. And the $w(\boldsymbol{p})$ also can be cross validated with numerical solving time-dependent Schr\"{o}dinger equation \cite{Zhou2017}. The QTM would provide more potential theoretical treatment for the TWG. For example, the long-range Coulombic potential can be introduced into QTM \cite{Yan2010} to explicitly investigate the influence of rescattering on the TWG.

The QTM directly correlates the TWGs and PEMDs: As well known, the PEMD results from the coherent sums of complex amplitude $\overset{\backsim}{M}_{\boldsymbol{p}}^{(\mathrm{SFA})}$ of quantum trajectories in the $\boldsymbol{p}$ space at $t=+\infty$; here, the TWG comes from the coherent sum of the photon wavepackets induced by electron quantum trajectories along $t$ but integrating over all trajectories in the $\boldsymbol{p}$ space. The QTM gives one-to-one correspondence between the TWGs and PEMDs when scanning $\tau$. The $w(\boldsymbol{p},\tau=0.34 \textrm{ fs})$ is most asymmetric when the TWG is maximum; the $w(\boldsymbol{p},\tau=0 \textrm{ fs})$ is symmetric when $|S_{\mathrm{THz}}|$ is minimum. The asymmetric electron distribution in the $\boldsymbol{p}$ space results in the most powerful TWG, which is equivalent to the explanation of the net current at $t=+\infty$ in the PC model. However, whether is the TWG related to the electron interference fringes in PEMDs? It is only can be understood with the recourse of the contributions of quantum trajectories.

\begin{figure*}[t]
  \includegraphics[width=0.9\linewidth]{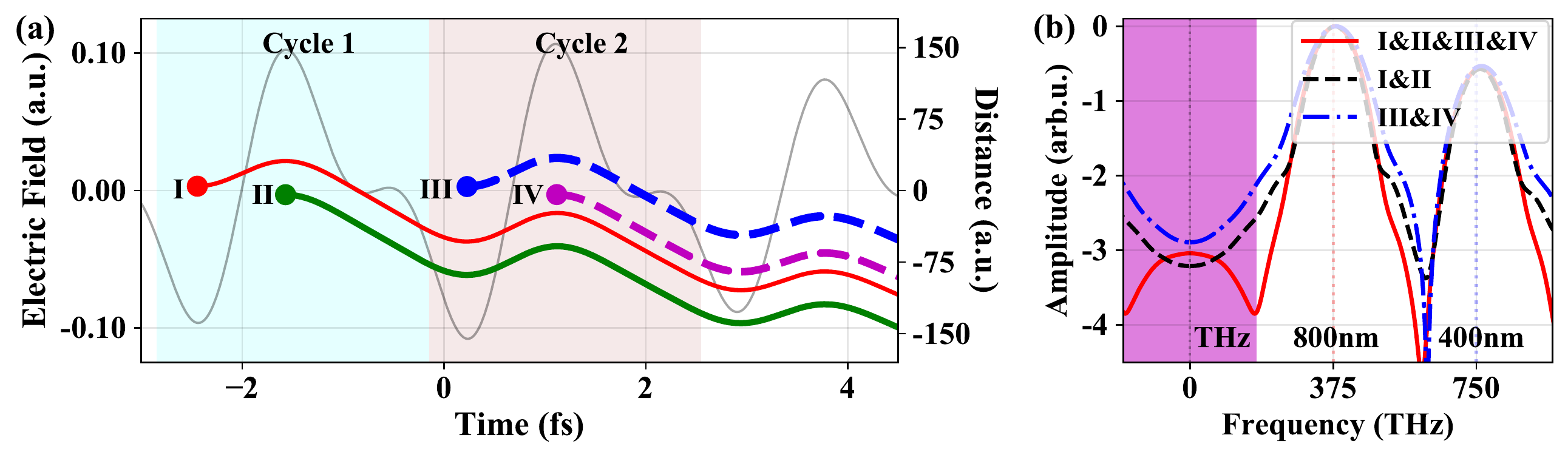}
  \caption{The backanalysis of the quantum trajectories and harmonic spectra. (a) The real-space quantum trajectories $\boldsymbol{r}_{t>t_{0}}^{(\alpha)}(t)$ (color lines) released at different ionization time ``slits'' of the dual-color electric field $E(t)$ (gray line). The different ionization ``slits'' are shaded with the cyan and magenta regions. (b) The harmonic spectra of the contributions from different combinations of electron trajectories. Only the interference of the trajectories from different ``slits'' produces the constructive THz peak. The harmonic spectra are logarithmically scaled and normalized to the maximum. }
\end{figure*}

We conduct the backanalysis by choosing the specific trajectories when $\tau=0.34 \textrm{ fs}$, and observing the Fourier components of $\boldsymbol{E}_{\mathrm{Har}}(t)$. The constraint condition is chosen such that the contributions of different types of quantum trajectories can be easily distinguishable. With the help of the backanalysis, we found that: (i) Any trajectory individually cannot produce the THz peak in frequency domain. The THz peak does only appear when a combination of trajectories are involved. (ii) Only few quantum trajectories are necessary for the convergence of the characteristics of harmonic spectra. The harmonic spectra only qualitatively change when more trajectories are involved. 

Then, the backanalysis method are used to clarify which kinds of trajectories are the most critical for the TWG. The real-space trajectories $\boldsymbol{r}_{t>t_{0}}^{(\alpha)}(t)$ at $\tau=0.34 \textrm{ fs}$ are chosen, and the analysis is reduced to the few characteristic trajectories with largest ionization possibilities, which are born at the peak of the electric field. The trajectories with the highest weights are plotted in Fig. 3 for demonstration of the trajectory backanalysis. The electron trajectories (color curves) are released at the different instants of the $\boldsymbol{E}(t)$ (gray curve). The thickness of the trajectories represents the relative weight $w^{(\alpha)}$ of the trajectory. The four trajectories are categorized into two types, depending on the tunnelling ``slits'': the ionization instants within one cycle or separately in neighbouring cycles. The trajectory I and II (solid trajectories) are released within cycle 1 (shaded as cyan color), while the trajectory III and IV (dash trajectories) are released in neighbouring cycles (magenta region). 

The photon wavepackets induced by the electron trajectories are calculated as Eq. \ref{eq:Harmonic radiation}. The various combinations of trajectories are coherently summed, and the Fourier spectra are examined, shown in Fig. 3(b). We find that only the interference between the trajectories born in the different cycles (the sum of solid and dash trajectories) results in the constructive THz peak in frequency component. The interference of the trajectories within the one cycle (solid trajectories or dash trajectories individually) cannot produce the peak structure in the THz region. The reason of the constructive interference is out-of-phase of the trajectories in successive cycles. The maximum peaks of the driving electric field, where the electron is most likely released, is split by the small secondary maxima, which slightly alter the phase accumulation of electron trajectories in successive cycles. The out-of-phase of the trajectories around the zero frequency induces the constructive interference in THz frequency region. Comparatively, the trajectories in the same cycle are phased together, so the constructive interference cannot appear. Here, the TWG mechanism is analogous with even-order high harmonics in dual-color field. In the HHGs, the electron trajectories released in successive half cycles recollide to emit the high-energy photon bursts in period of half cycle. The perturbative $2\omega$ field unbalances the trajectories and breaks the symmetry of electron recollision, thus resulting in the constructive interference of even harmonics. But they are not completely same. The appearance of even harmonics considers the recollision electron, but the TWG concerns direct electron. In even harmonics generation, the weak $2\omega$ field breaks the symmetry of electron trajectories in successive half cycles; but in the TWG, the $2\omega$ field breaks the phase synchronization of trajectories in the neighbouring cycles.

The trajectory analysis provides a completely new and intuitive understanding of the TWG in strong field laser from the semiclassical theory. The portion of the valence electron is released at the different instant $t_{\mathrm{ion}}$ with the ionization possibility $w(t_{\mathrm{ion}})$. The subsequent motion of the free electron is treated classically, described as quantum trajectoris $\boldsymbol{r}_{t>t_{\mathrm{ion}}}(t)$, and the photon emission originates from the acceleration of the free electron in the rapidly oscillatory strong laser, according to the Maxwell theory. Only the interference of the electron trajectories appearing in the neighboring cycles, e.g. intra-cycle interference, leads to the constructive peak in THz frequency region; the interference within the cycles (inter-cycle interference) cannot produce the peak structure in the THz region.

\begin{figure}[h]
  \includegraphics[width=\linewidth]{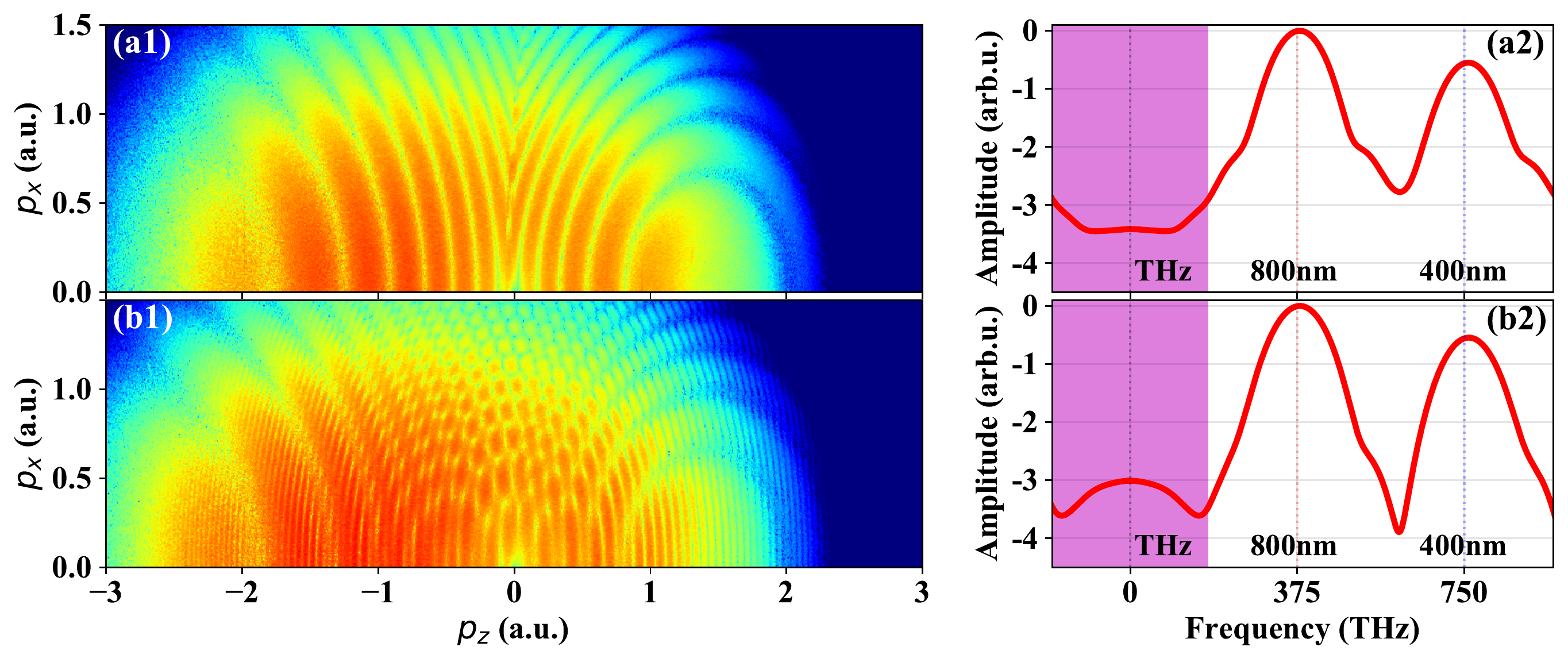}
  \caption{The link between the photoelectron spectra and harmonic spectra. (a1) and (a2) The photoelectron and harmonic spectra reconstructed by the trajectories born at the tunnelling ``slit'' of Cycle 1, (cyan region in Fig. 3(a1)). (b1) and (b2) The photoelectron and harmonic spectra reconstructed by the trajectories born at the Cycle 1 and Cycle 2 (cyan and magenta regions). All the values are logarithmically scaled and normalized to the maximum.}
\end{figure}

The QTM establishes a straightforward link between the PEMDs $w(\boldsymbol{p},\tau=0.34 \textrm{ fs})$ and Fourier spectra of $\boldsymbol{E}_{Har}(t)$. The constraint condition in the time domain is chosen to filter out the quantum trajectories with different ionization instants, and the PEMDs and harmonic spectra are plotted in Fig. 4. The PEMD and harmonic spectrum reconstructed by the trajectories within the same cycle, marked as ``Cycle 1'' (cyan region in Fig. 3(a)), are shown in Fig. 4(a). The divergence pattern of ``wings'' fringes in Fig. 4(a1) is the characteristic feature of inter-cycle interference. Although the asymmetry of the PEMD still exists, the TWG cannot be produced, which is in contradication with the explanation of the PC model. The trajectories from ``Cycle 2'' (\textit{Supplemental Material S.3}) cannot produce the TWG as well. The contribution of the trajectories, released in both ``Cycle 1'' and ``Cycle 2'', is shown in Fig. 4(b). Compared with the ``wings'' fringes, additional ``rings'' fringes appear on the PEMD due to the intra-cycle interference, which results in the constructive THz peak. The link between the PEMDs and TWGs confirms that the intra-cycle interference results in the TWG and simultaneously the ``rings'' fringes in photoelectron spectra, whereas the inter-cycle interference does not create the TWG. Our analysis exhibits that the asymmetric PEMD, corresponding to the net current in PC model, is necessary but not sufficient for the TWG, while the intra-cycle interference is necessarily demanded. Our supposition can be examined by exploiting the pulse compression technique. The TWG should disappear when the dual-color field is compressed to single-cycle regime \cite{Hassan2016}, since the intra-cycle interference is suppressed due to closed ionization slits.

In this letter, the semiclassical method, so called quantum trajectories, is developed to explain the terahertz generation in the strong dual-color laser, and the delay-dependent TWGs calculated with QTM is verified with the experiment. Our investigation exhibits that the TWG originates from the quantum path interference of electron trajectories released from neighboring cycles of the dual-color electric field, whereas the interference of trajectories within one cycle cannot produce TWG. Here, the TWG in the dual-color field  is treated as a new-type electron tunnelling interferometry, which has manifold meanings. Firstly, through artificial design of the driving electric field with the pulse shaping technique, the electron paths in tunnelling interferometry can be controlled for efficient TWG for further applications \cite{Zhang2018}. Secondly, the TWG in dual-color field would act as an all-optical, ultrafast, vacuum-free tunnelling interferometry to probe the electron dynamics in gas \cite{Pedatzur2015} and solid \cite{Vampa2015} mediums.

National Natural Science Foundation of China (NSFC) (11827806, 61675213, 11874368). We also acknowledge the support from Shanghai-XFEL beamline project (SBP) and Shanghai High repetition rate XFEL and Extreme light facility (SHINE).

Y. Z. and K. Z. contributed equally to this work.

\bibliography{main}

\appendix

\section{Quantum Trajectroy Method (QTM)}

The terahertz (THz) wave generations (TWGs) and photoelectron momentum distributions (PEMDs) can be calculated with the QTM. In the calculation, the fundamental $\omega$ and second harmonic $2\omega$ electric fields are written as the form of $\boldsymbol{A}(t)=-\frac{E}{\omega}\boldsymbol{e}_{z}\sin^{2}(\frac{\omega t}{2n})\sin(\omega t)$ for $t\in [0,\frac{2n\pi}{\omega}]$. The $\omega$ pulse consists of 8 cycles, and the $2\omega$ pulse has the same duration of 16 cycles. The peak field strength $E_{\omega}=0.075 \text{ a.u.}$ and $E_{2\omega}=0.045 \text{ a.u.}$, and the fundamental frequency $\omega=0.05695$. The atom system is the 1$s$ state of hydrogen atom. The large sampling N is necessary to obtain the subtle structure of the PEMD, but only few trajectories can reproduce the \textit{sin}-type TWG yields as a function of phase delay $\tau$ of the dual-color laser. Here, the THz peak-peak amplitudes $S_{\mathrm{THz}}(\tau)$ is obtained by the integral of ~3000 trajectories randomly sampled in $\boldsymbol{p}$ space. The convergence were cautiously examined. 

\section{Experiment}

The experimental setup is shown in Fig. S1. The femtosecond laser (Libra, Coherent Inc.) delivers a femtosecond pulse (800 nm, $\sim$35 fs, $\sim$1.75 mJ), which was guided into the experimental setup. A beam splitter separated the pulse into two parts with 96\% and 4\% pulse energy. The 96\% beam was used as the pump beam for the TWG, and the 4\% beam was employed as the probe beam for time-domain terahertz electro-optic sampling (EOS) method. 

The pump beam passed through a 200-$\mathrm{\mu m}$ type-I BBO crystal with the double-frequency efficiency of $\sim$23\%. The fundamental beam $\omega$ was $p$-polarized, and the optical axis of BBO was always kept perpendicular with the $\omega$ polarization to obtain the maximum up-conversion efficiency. The $s$-polarized $2\omega$ beam propagated collinearly with the $\omega$ beam. Because of different refractive indices of $\omega$ and 2$\omega$ in air, the relative phase delay $\tau$ of $\omega$-$2\omega$ pulses can be precisely controlled by moving the BBO crystal along the propagation direction. The relative polarization angle $\theta$ between the $\omega$ and $2\omega$ beams can be rotated by a zero-order dual-wavelength wave plate (DWP), which acts as a half-wave plate for the $\omega$ beam and a full-wave plate for the $2\omega$ beam. The polarization of the $2\omega$ beam was aligned parallel with the $s$-polarized $\omega$ beam, that $\theta=0^{\circ}$ throughout our measurement. Due to the collinear geometry of the $\omega$-$2\omega$ beam, the $\tau$ between $\omega$-2$\omega$ pulses can be passively stabilized up to sub-wavelength accuracy.

The dual-color laser pulses were focused by a silver parabolic mirror with an effective focal length of $\sim$ 100 mm to ionize air for the TWG. Here, we use a tightly-focusing scheme to deliberately prevent the propagation effect in plasma. The TWG was collected and collimated with a gold parabolic mirror with 100-mm focal length, and focused into 1-mm-thick (110)-cut ZnTe crystal with a same parabolic mirror. A 500-$\mu m$-thick polished silicon wafer reflected the residual laser, only allowing for the THz transmission. A pellicle beam splitter combined the THz pulse and the weak probe beam to implement the free-space EOS detection. The terahertz electric field $E_{\rm{THz}}(t)$ has a signal-to-noise ratio (SNR) better than 100:1. A metal wire-grid THz polarizer filtered out the $s$-polarized components of the TWG. The $S_{\rm{THz}}(\tau)$, shown in Fig. 2(a1) in the main text, is defined as $S_{\rm{THz}}=\pm \left| \rm{max}[E_{\rm{THz}}(t)]- \rm{min}[E_{\rm{THz}}(t)]\right|$, and the sign $\pm$ represents the polarity of the THz waveform. 

\begin{figure}[h]
  \includegraphics[scale=1]{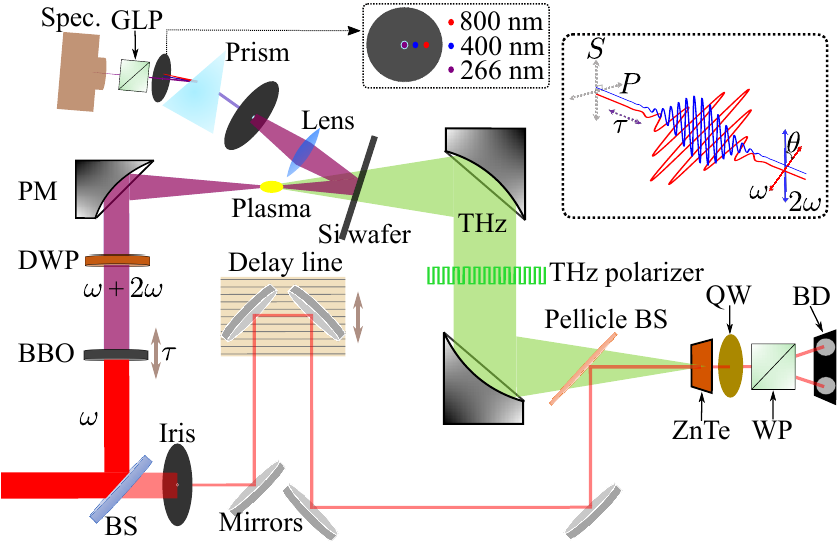}
  \caption{Experimental setup. BS: beam splitter; CH: chopper; $\beta$-BBO: beta barium borate; DWP: dual-wavelength plate; PM: parabolic mirror; QWP: quarter wave plate; GLP: glan-laser polarizer; WP: wollaston polarizer; BD: balanced detector.}
\end{figure}

The third harmonics generation (THGs) of $\sim$ 266 nm was jointly measured with the TWG for the \textit{in situ} calibration for $\tau$ of the $\omega$-$2\omega$ beam. The THG was reflected by the silicon wafer with residual $\omega$ and $2\omega$ beams, and spectrally separated by a suprasil prism. The $s-$components of THGs $S_{3\omega}(\tau)$ were decomposed with a glan-laser polarizer, and collected into a fiber spectrometer, shown in Fig.2(a2) in the main text. The THGs were measured with 50-ms integration time, 10-time average in our measurement. Exploiting with the $\tau$-dependent THGs and simulations, the $\tau$ can be calibrated up to the sub-wavelength accuracy. 

\section{The contribution of partial trajectories}

The trajectories released at different tunnelling slits are chosen to reconstruct the PEMDs and harmonic generation spectra. The PEMDs and TWGs, reconstructed with the trajectories within the ``Cycle 1'' and ``Cycle 1+Cycle 2'' have been shown in Fig. 4 in the main text. The PEMD and harmonic spectrum, reconstructed with the trajectories within the ``Cycle 2'', are shown in Fig. S2 in \textit{supplemental material} for comparison. The trajectories from individual ``Cycle 1'' and ``Cycle 2'' only produce the inter-cycle interference with ``wings'' fringes and no constructive peak in THz region. The trajectories from ``Cycle 1+Cycle 2'' result in the  intra-cycle interference with ``ring'' fringes and TWGs.

\begin{figure}[h]
  \includegraphics[width=\linewidth]{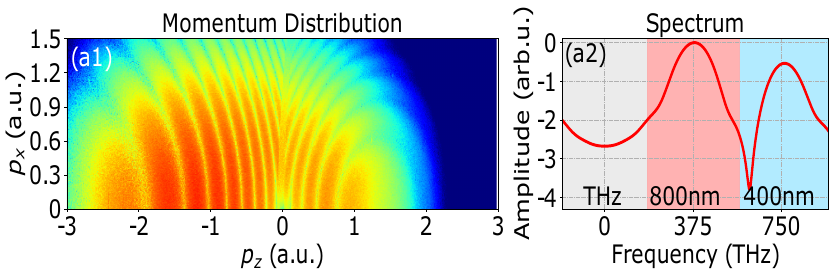}
  \caption{The PEMDs (a1) and harmonic spectra (a2) due to the trajectories released in the ``Cycle 2'' region. }
\end{figure}

\end{document}